# Relative Citation Ratio (RCR):

# An empirical attempt to study a new field-normalized bibliometric indicator

Lutz Bornmann* & Robin Haunschild**

*Division for Science and Innovation Studies

Administrative Headquarters of the Max Planck Society

Hofgartenstr. 8,

80539 Munich, Germany.

Email: bornmann@gv.mpg.de

**Max Planck Institute for Solid State Research

Heisenbergstr. 1,

70569 Stuttgart, Germany.

Email: R.Haunschild@fkf.mpg.de

**Abstract**

Hutchins, Yuan, M., and Santangelo (2015) proposed the Relative Citation Ratio (RCR) as a new field-normalized impact indicator. This study investigates the RCR by correlating it on the level of single publications with established field-normalized indicators and assessments of the publications by peers. We find that the RCR correlates highly with established field-normalized indicators, but the correlation between RCR and peer assessments is only low to medium.

**Key words**

Relative Citation Ratio; RCR; bibliometrics; field-normalization

# 1 Introduction

It is standard in bibliometrics to field-normalize citation counts (Vinkler, 2010). For cited-side normalization, the citation counts of a focal paper are compared with the citation counts of a reference set: all papers published in the same field and year. The fields are defined on the basis of journal sets, subject categories, which are algorithmically constructed on the basis of citation relations between publications (Ruiz-Castillo & Waltman, 2015), or expert opinions. For citing-side normalization, each citation of a focal paper is weighted by the citation density of the citing paper's field (Bornmann & Marx, 2015). Hutchins et al. (2015) – a team of authors affiliated to the US National Institutes of Health (NIH) – proposed the Relative Citation Ratio (RCR) as a new field-normalized impact indicator. This indicator could signify an interesting new approach to normalization on the cited-side, because it relies on co-citations to generate the reference set. The papers co-cited with the focal paper are considered to represent the field of the paper. In bibliometrics, co-citation is a similarity measure for papers which is based on citation relationships (another measure is bibliographic coupling).

This study investigates the correlation of RCR on the level of single publications with established field-normalized indicators and assessments of publications by peers.

# 2 Literature overview and alternative field-normalized indicators

The RCR has received positive comments because it could mean a move away from the use of the Journal Impact Factor (JIF) in the biomedical area: Stefano Bertuzzi, executive director of the American Society for Cell Biology in Bethesda, "applauds the NIH for moving away from the journal impact factor (JIF). He wrote that the metric 'evaluates science by putting discoveries into a meaningful context. I believe that the RCR is a road out of the JIF swamp'" (Bloudoff-Indelicato, 2015). The JIF measures the average citation impact of journals without normalizing citations and is frequently misused (in biomedicine) to study the impact of single publications. Many other moves away from the JIF were made, see for example the San Francisco Declaration on Research Assessment (DORA, http://www.ascb.org/dora/). Waltman (2015) criticizes the RCR using a fictitious example publication receiving citation impact from different fields. He shows that "publications may be penalized rather than rewarded for receiving interdisciplinary citations". The receiving of new citations (from a discipline with high citation density) could mean that the RCR of a publication is decreasing instead of increasing. Waltman (2015) regards this property of the RCR as an important disadvantage which disqualifies it from being an equitable alternative to the field-normalized indicators already in use in bibliometrics. The following three established indicators are currently in use in bibliometrics and reflect different field-normalizing methods (Bornmann & Marx, 2015):

(1) The Mean Normalized Citation Score (MNCS) is a quotient composed of the citation counts of a focal paper (numerator) and the average citation counts in the reference set (denominator, see above) (Waltman, van Eck, van Leeuwen, Visser, & van Raan, 2011a; Waltman, van Eck, van Leeuwen, Visser, & van Raan, 2011b). (2) For the calculation of Citation Percentiles (CP), the papers in the reference set are sorted by citation counts in descending order and the proportion of papers with equal or lower citation counts is calculated for every paper (and also the focal paper) (Bornmann, Leydesdorff, & Mutz, 2013). (3) MNCS and CP are cited-side normalization approaches. For the $SNCS_{(2)}$ which is an indicator on the basis of citing-side normalization, each citation to a paper is weighted with the number of cited references in the citing paper (Waltman & van Eck, 2013). The idea behind this indicator is that the number of references reflects the citation density of the field in which the citing paper was published. All indicators used in in this study are described in detail by Bornmann and Marx (2015).

# 3 Data set

F1000Prime is a post-publication peer review system of papers from the medical and biological areas (see http://f1000.com/prime). The papers included in the system are rated by Faculty members (leading scientists and clinicians) as "Good", "Very good", or "Exceptional" which are equivalent to scores of 1, 2, or 3, respectively. In many cases a paper is not evaluated by one Faculty member alone but by several. F1000Prime provided one of the authors of this study with data on all ratings made and the bibliographic information for the corresponding papers in their system (Bornmann, 2014, 2015a, 2015b; Bornmann & Haunschild, 2015). Since the dataset does not contain any citation impact scores, it was matched with a bibliometric in-house database at the Max Planck Society (MPG), which is administered by the Max Planck Digital Library (MPDL) and is based on the Web of Science (WoS, Thomson Reuters). To enable a citation window of at least three years for every publication (Glänzel, 2008), those published later than 2012 are discarded. Thus, publications published between 1996 and 2012 are included.

For this study, the F1000Prime dataset is reduced to only those publications with at least two Faculty members' scores each. In order to increase the reliability of the scores, publications with only one score are excluded. Assessments by experts might be personally biased and the consideration of more than one score should increase the reliability of the assessments (Bornmann & Marx, 2014). In order to have only one total score from the Faculty members, an average is calculated over the members' scores for one and the same paper. The consideration of only those papers published before 2013 with (1) at least two scores of Faculty members and (2) bibliometric data available in the in-house database reduces the F1000Prime dataset to n=16,557 papers for this study. For n=16,521 publications an RCR score could be retrieved from https://icite.od.nih.gov/ on November 10, 2015.

# 4    Results

Table 1 shows the Spearman rank-order correlation coefficients for the correlations between MNCS, CP, SNCS$_{(2)}$, F1000 score, citations and RCR. Additionally, Figure **1** shows the Scatterplot matrix among the variables in order to visualize the relationships. We included for every paper the number of citations for a three year citation window (starting one year subsequent to the publication year) besides the normalized indicators. Whereas the correlation between RCR, citations and F1000 score is based on all publications in the set of this study (n=16,521), the correlations with the MNCS and CP are based on a slightly reduced set with n=16,518 publications (for three publications, the indicators are not both available). Since the SNCS$_{(2)}$ is only available for the years 2007 to 2010 in the MPG in-house database, the correlations with the indicator are based on n=7,055 publications. The interpretation of the correlation coefficients follows the guideline of Cohen (1988).

Table 1. Spearman rank-order correlation coefficients for the correlations between MNCS, CP, SNCS$_{(2)}$, Citations, F1000 score, and RCR

|  | RCR | MNCS | CP | SNCS$_{(2)}$ | Citations | F1000 score |
|---|---|---|---|---|---|---|
| RCR | 1.00 | | | | | |
| MNCS | 0.88 | 1.00 | | | | |
| CP | 0.85 | 0.99 | 1.00 | | | |
| SNCS$_{(2)}$ | 0.85 | 0.78 | 0.75 | 1.00 | | |
| Citations | 0.70 | 0.64 | 0.61 | 0.88 | 1.00 | |
| F1000 score | 0.29 | 0.25 | 0.23 | 0.31 | 0.25 | 1.00 |

Note. CP is calculated on the basis of the formula proposed by Hazen (1914).

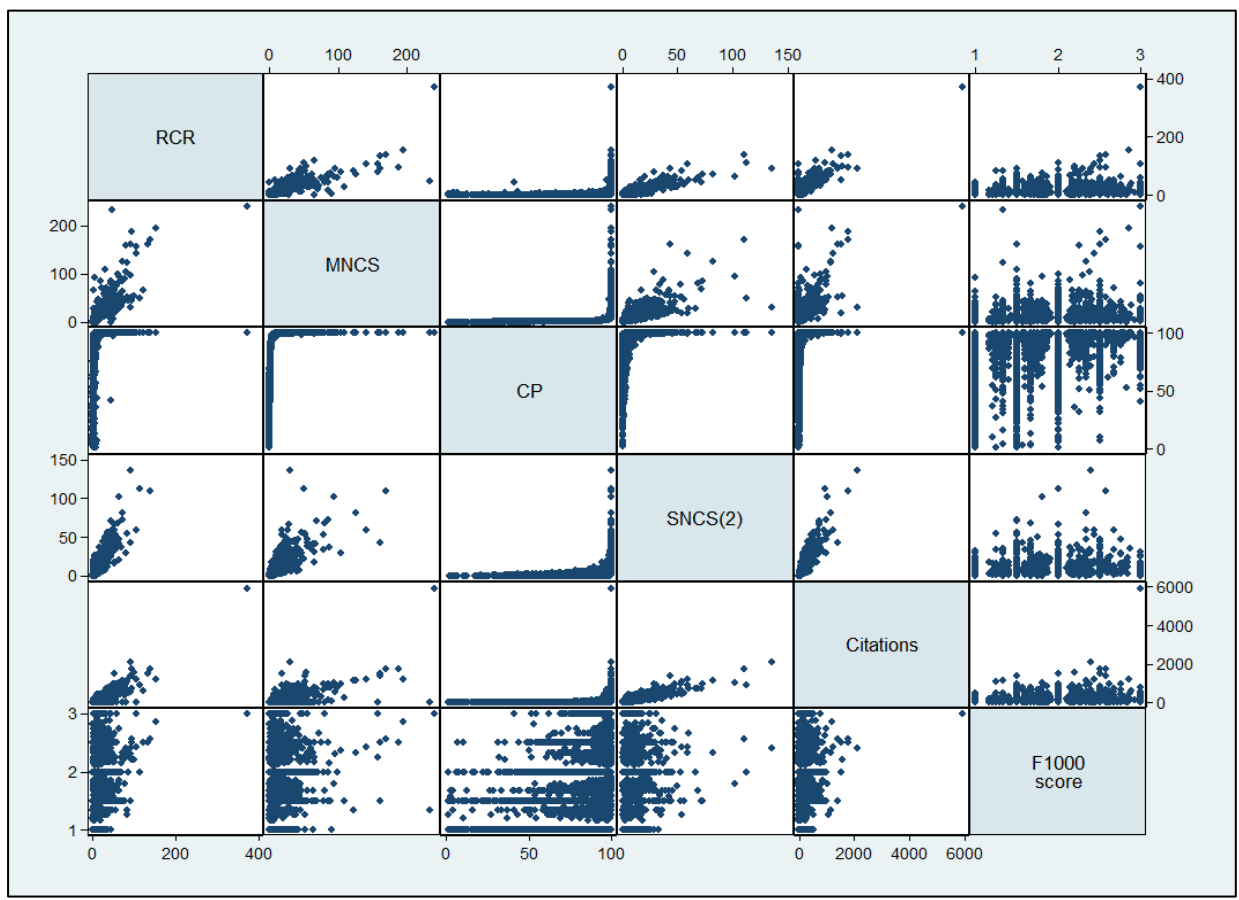

Figure 1. Scatterplot matrix among MNCS, CP, SNCS$_{(2)}$, Citations, F1000 score, and RCR

As Table 1 shows, the RCR correlates highly with all other field-normalized indicators (MNCS, CP, and SNCS$_{(2)}$). The correlation between RCR and citations is somewhat lower, but can be also interpreted as high. When interpreting the coefficients, one should consider that Hutchins et al. (2015) only included biomedical citations because the calculation of the RCR is based on the PubMed database (i.e. PubMed-to-PubMed citations). Any citations from papers not covered by the PubMed database are not counted. The other indicators in Table 1 are based on WoS data which is a multi-disciplinary database including citations from a broad range of disciplines.

The correlations between the F1000 score and the bibliometric indicators are on a significantly lower level. However, the lower coefficients are in agreement with the results of other studies which correlated F1000 scores and citation metrics: The meta-analysis of Bornmann (2015a) reveals a pooled r=0.25. The pooled coefficient and the coefficients in Table 1 can be interpreted as between low and medium, whereas CP correlates the lowest ($r_s$=0.23) and SNCS$_{(2)}$ the highest ($r_s$=0.31) with the F1000 scores. There are very slight differences between the coefficients of the normalized indicators and citations.

In this study, we have used the average value of the F1000 recommendation scores for each paper. Another approach pursued by Hutchins et al. (2015) is to use the sum over all F1000 scores for each paper. Using the sum over all F1000 scores for each paper results in slightly increased correlation coefficients: They range between $r_s$=0.27 for CP and $r_s$=0.38 for SNCS$_{(2)}$, but do not differ significantly from the correlation coefficients in Table 1 where the average value over all F1000 scores per paper is used.

## 5   Discussion

Our results show that RCR correlates highest with MNCS, closely followed by CP and SNCS$_{(2)}$. This is surprising considering the valid criticism by Waltman (2015). Probably, the fictitious example in Waltman (2015) does not occur often in the publication set studied here. Papers recommended by F1000 are usually from the biomedical fields and usually receive their citations from there, so that the fictitious example in Waltman (2015) does not occur in many cases within the F1000Prime publication set.

Hutchins et al. (2015) also studied the correlation of RCR with F1000 scores and reported a correlation coefficient of r=.44 (in the supplemental information). This coefficient is somewhat higher than the one calculated here (r=0.29). However, their validation was done on a smaller (n=2,193 papers) and more selective publication set (R01-funded papers published in 2009). Our study includes a larger publication set (n=16,521 papers) with a range of publication years (from 1996 to 2012) and imposes no funding restrictions at all.

Even larger scale studies are desirable when RCR scores are also available for publications outside the biomedical area. The database PubMed focuses on this area and one needs PubMed IDs to retrieve RCR scores. If the RCR scores were available for publications covering more areas (e.g., publications without a PubMed ID), it could be investigated whether the RCR can field-normalize citation counts better than established field-normalized indicators in bibliometrics. Furthermore, it would be necessary for a larger study that RCR scores be obtained for larger publication sets (without the restriction of processing 200 papers at a time, see https://icite.od.nih.gov/). It would be very helpful if an application programming interface (API) were provided by the NIH for the purpose of comparing RCR values of a large amount of papers with the field-normalized indicators currently in use in bibliometrics. There are various methods available which can be used to study the ability of the RCR to normalize citation counts (Bornmann, de Moya Anegón, & Mutz, 2013; Waltman & van Eck, 2013). These methods should be used to compare the RCR with established field-normalized indicators.

# Acknowledgements

We would like to thank Ros Dignon and Iain Hrynaszkiewicz from F1000 for providing one of the authors with the F1000Prime data set. The bibliometric data used in this paper is from an in-house database developed and maintained by the Max Planck Digital Library (MPDL, Munich) and derived from the Science Citation Index Expanded (SCI-E), Social Sciences Citation Index (SSCI), Arts and Humanities Citation Index (AHCI) provided by Thomson Reuters. We would like to thank Ludo Waltman and Ian Hutchins for their valuable feedback on an earlier version of our paper.